\documentstyle[12pt]{article}
\def\be{\begin{equation}}
\def\ee{\end{equation}}
\def\bea{\begin{eqnarray}}
\def\eea{\end{eqnarray}}
\def\a{\alpha}
\def\b{\beta}

\def\pa{\partial}
\def\e{\epsilon}
\def\d{\delta}

\author{Hans-J\"urgen Schmidt}

\title{The metric in the superspace of Riemannian
metrics and its  relation to gravity\footnote{This paper is in final form 
and no version of  it will be submitted for publication elsewhere.} }

\date{}
\begin{document}
\maketitle

\centerline{Universit\"at Potsdam, Institut f\"ur Mathematik, Am
Neuen Palais 10} 
 \centerline{D-14469~Potsdam, Germany,  E-mail:
 hjschmi@rz.uni-potsdam.de}

\begin{abstract}
\noindent
The space of all Riemannian metrics is  infinite-dimensional. Nevertheless 
a great deal 
of usual Riemannian geometry can be carried over. The superspace of all
Riemannian 
metrics shall be endowed with a class of Riemannian metrics;
 their curvature and 
invariance properties are  discussed. Just one of this class
 has the property to bring the
 lagrangian of General Relativity into the form of a classical particle's
motion. 
The signature of the superspace metric depends in a non-trivial manner
 on the signature 
of  the 
original metric, we derive the corresponding formula. Our approach is  a
local 
one: the essence  is a metric in the space of all symmetric rank-two tensors,
 and then 
 the space becomes a warped product of the real line with an Einstein space.


\noindent
MS classification: 53C20
\end{abstract}

\section{THE SUPERSPACE}

Let $n \ge 2$, $n$ be the dimension of   the basic 
 Riemannian spaces. Let $M$ be an 
$n$-dimensional differentiable manifold with an 
atlas $x$ of coordinates $x^i$, 
 $i = 1, \dots, n$. The signature $s$ ( $=$ number of negative 
eigenvalues) shall be fixed; let $V$  be the space of all Riemannian 
metrics $g_{ij}(x)$  in $M$  with signature $s$, related to the coordinates
$x^i$. 
This implies that isometrical metrics in $M$ are different points in $V$
in general. 
The $V$ is called superspace,  its points are the Riemannian metrics. The
tangent 
space in $V$  is the vector space
\be
     T = \{ h_{ij}(x) \vert   x \in M, \  h_{ij}= h_{ji} \}
\ee
the space  of all symmetric 
tensor fields of rank 2. 
All considerations are local ones, so we may have in mind 
one single fixed coordinate system in $M$.

\section{COORDINATES IN SUPERSPACE}

Coordinates should possess one contravariant index, 
so we need a transformation of the type
\be
 y^A     = \mu^{Aij} g_{ij}(x)
\ee
such that the $y^A$  are the coordinates for $V$.
To have a defined one--to--one correspondence
 between the index pairs $ (i, \, j)$ 
 and the index $A$
 we require 
$$
A = 1, \dots N = n(n + 1)/2 \, ,
$$
 and $A = 1, \dots  N$ 
 corresponds 
to the pairs
\bea
(1,1), \ (2,2),  \dots (n,n), \ (1,2), \ (2,3),
 \dots (n-1,n), \ (1,3),   \nonumber \\ 
  \dots (n-2,n), \dots (1,n)
\eea 
consecutively. $(i, \, j)$ and 
$(j, \, i)$  correspond to the same $A$. We make the ansatz
\bea
    \mu^{Aij} = \mu_{Aij} = 
 b  \ {\rm for} \  i \ne j, \quad  
 c \ {\rm for} \  i = j \nonumber  \\
          0 \ {\rm if} \  (i,j) \ {\rm does \ not \ correspond \  to \ } A   
\eea
and require the usual inversion relations
\be
 \mu^{Aij} \mu_{Bij} = \d^A_B
\quad {\rm  and}  \quad 
\mu^{Aij}\mu_{Akl}
= \d_k^{(i} \d_l^{j)} \, . 
\ee
Bracketed indices are to be symmetrized, 
which is necessary because of symmetry 
of the metric $g_{ij}$.
 Inserting ansatz (4) into (5) gives $c^2 = 1$,
 $b^2 = 1/2$. Changing 
the sign of $b$ or $c$ only changes the sign 
of the coordinates, so we may put
\be
     c=1, \qquad  b=1/\sqrt 2   \, .
\ee
The object $ \mu^{Aij}$ is
 analogous to the Pauli spin matrices relating two 
spinorial indices to one vector index.

\section{METRIC IN SUPERSPACE}

The metric in the superspace shall be denoted by $H_{AB}$,
 it holds
\be
    H_{AB} = H_{BA}
\ee
and the transformed 
metric is
\be
G^{ijkl} = H_{AB}     \mu^{Aij}    \mu^{Bkl} \, , \quad
H_{AB} =      \mu_{Aij}    \mu_{Bkl}
G^{ijkl} \, .
\ee
From (4) and (7) it follows that
\be
G^{ijkl} = G^{jikl} = G^{klij} \, .
\ee
The inverse to $H_{AB}$  is $H^{AB}$, and we define
\be
G_{ijkl} = H^{AB}     \mu_{Aij}    \mu_{Bkl}
\ee
which has the same symmetries as (9). We
 require 
$G^{ijkl}$
 to be a tensor and use only the metric $g_{ij}(x)$  to define it. 
Then the ansatz
\bea
G^{ijkl}
= z \,  g^{i(k} \,  g^{l)j} + \a \,  g^{ij} \,  g^{kl}
\\
 G_{ijkl}    = v \, g_{i(k} \,  g_{l)j} + \b \,  g_{ij} \,  g_{kl}
\eea
where $v$, $z$, $\a$ and $\b$ are constants,  is 
the most general one to fulfil  the symmetries (9). One 
should mention that also curvature-dependent constants 
could have heen introduced. The requirement 
that $H_{AB}$  is the inverse to $H^{AB}$  leads via (8,10) to
\be
G_{ijkl} G^{klmp}
 = 
 \d_i^{(m} \d_j^{p)} \, . 
\ee

The requirement that  $G^{ijkl}$ 
is a tensor can be 
justified as follows: Let 
a curve $y^A(t)$, $0 < t < 1$  in $V$ be given, 
then its length is 
$$
\sigma  = \int_0^1 \left(
 H_{AB} 
\frac{dy^A }{dt}\frac{dy^B }{dt}
 \right)^{1/2} \, dt
$$
i.e., 
with (2) 
and (8)
\be
\sigma  = \int_0^1 \left(
G^{ijkl}  
\frac{d g_{ij} }{dt}\frac{d g_{kl} }{dt}
 \right)^{1/2} \, dt    \, .
\ee
A coordinate transformation in $M: x^i \to \epsilon x^i$ 
changes   $g_{ij} \to \epsilon^{-2} g_{ij}$.
 We now require that $\sigma$ shall not 
be changed by such a transformation. 
Then $\a$ and $z$ are constant real numbers. Inserting 
(11,12) into (13) gives $v \, z = 1$, hence $z \ne 0$. 
By a  constant rescaling we get
\be
v = z = 1
\ee
and then (11,12,13) yield
\be
\a \ne \frac{-1}{n} \, , \quad \b = \frac{- \a}{1 + \a n}   \, .
\ee
So we have got a one-parameter 
set of metrics in $V$. Eq. (16) fulfils the 
following duality relation: with 
$f(\a) = - \a/(1 + \a n)$,
 $f(f(\a)) = \a$ holds for all $\a \ne -1/n$.
\bea
G^{ijkl}
=   g^{i(k} \,  g^{l)j} + \a \,  g^{ij} \,  g^{kl}
\nonumber  \\
 G_{ijkl}    =  g_{i(k} \,  g_{l)j} + \b \,  g_{ij} \,  g_{kl} \,.
\eea
It holds: For $\a = -1/n$, the metric $H_{AB}$  is not invertible.

\noindent
{\bf Indirect proof:} $G^{ijkl}$
 depends continuously on $\a$, so it must 
be the case with the inverse. 
But 
$$
\lim_{\a \to -1/n}
$$
 applied to $ G_{ijkl}    $
 gives no finite result. Contradiction.

\section{Signature of the superspace metric}

Let $S$ be the signature 
of the superspace metric $H_{AB}$.
 $S$ depends on $\a$ and $s$.
 For convenience we define
\be 
\Theta =
  0, \quad ( \a>-1/n) \qquad
 1, \quad ( \a<-1/n) \, .
\ee
From continuity reasons it follows 
that $S$  is a function of $\Theta$ and $s$: $S = S(\Theta, s)$. 
If we transform 
$g_{ij} \to - g_{ij}$
 i.e., $s \to  n - s$, then $H_{AB}$  is not changed, i.e.,
\be
     S(\Theta,s) = S(\Theta,n - s) \, .
\ee
We transform $g_{ij}$  to diagonal form as follows
\be
g_{11} = g_{22} = \dots = g_{ss} 
 = -1, \quad g_{ij} = \d_{ij} \quad {\rm otherwise}\, .
\ee

\subsection{Signature for $\Theta = 0$}

 To calculate $S(0,s)$ we may put $\a = 0$  and get with (8,11,15)
\be
    H_{AB} =      \mu_{Aij} \,   \mu_{Bkl} \, 
g^{ik} \,  g^{jl} 
\ee
which is a diagonal matrix. 
It holds $H_{11} = \dots   = H_{nn}    = 1$  and 
the other diagonal components are $\pm 1$. 
A full estimate gives in agreement with (19)
\be
     S(0,s) = s(n - s) \, . 
\ee

\subsection{Signature for $\Theta = 1$}
 To calculate $S(1,s)$  we may put $\a = -1$ and get
\be
    H_{AB} =      \mu_{Aij} \,   \mu_{Bkl} \, \left(
g^{ik} \,  g^{jl}  -   g^{ij} \,  g^{kl}    \right) \, .
\ee
For $A \le  n < B$, $H_{AB} = 0$, i.e., 
the matrix $H_{AB}$  is composed of two blocks. 
For $A, \, B \le  n$ we 
get
$$
H_{AB} =      
  0 \quad {\rm for}   \quad  A=B, \qquad
 1    \quad {\rm for}   \quad  A \ne B
$$
a matrix which has the $(n - 1)$-fold
 eigenvalue 1 and the single eigenvalue $1 - n$.
For $A,B > n$ we have the same
 result as for the case $\a = 0$, i.e., we get
 $ S(1,s) = 1 + s(n - s)$.

\subsection{Result}
 The signature of the superspace metric is
\be
S = \Theta  + s(n - s) \, .
\ee

\section{SUPERCURVATURE}

We use exactly the same formulae  
as for finite-dimensional Riemannian 
geometry to define Christoffel affinities 
$\Gamma^A_{BC}$
 and 
Riemann tensor $R^A_{BCD}$.  Using (4) we write all
equations with indices $i,j=   1,\dots n$. Then each pair 
of covariant indices $i,j$ corresponds
 to one contravariant index $A$. The following formulae appear:
\bea
\frac{\partial g^{ij}}{\pa g_{km}} = - g^{i(k} \, g^{m)j}
\\
\Gamma^{ijklmp} = - \frac{1}{2}g^{i(k} g^{l)(m} g^{p)j}
- \a g^{ij} g^{k(m} g^{p)l}- \frac{1}{2}
 g^{j(k} g^{l)(m} g^{p)i}
\eea
and, surprisingly independent of $\a$  we get
\be
\Gamma^{klmp}_{ij}
=      -
   \d^{(k}_{(i} \, g^{l)(m}
 \, \d^{p)}_{j)} \, .
\ee
Consequently, also Riemann- and Ricci tensor do not depend 
on $\a$:
\be
R_{rs}^{klmpij} = \frac{1}{2} \left(
   \d^{(k}_{(r} \,g^{l)(m}g^{p)(i}
 \, \d^{j)}_{s)}
-
   \d^{(k}_{(r} \,g^{l)(i}g^{j)(m}
 \, \d^{p)}_{s)}
\right) \, .
\ee
Summing over $r=  m$ and $s = p$ we get
\be
R^{klij}
= \frac{1}{4} \left(
g^{ij} g^{kl} - n g^{k(i} g^{j)l}
\right) \, .
\ee
The Ricci tensor has one eigenvalue 0. 
Proof: It is not invertible because it 
is proportional to the metric  for the degenerated case 
$\a = -1/n$, cf.  sct. 3.

\medskip

The co--contravariant Ricci tensor reads
\be
R^{ij}_{kl} = G_{klmp} R^{mpij} 
= \frac{1}{4} \left(
g^{ij} g_{kl} - n \d_k^{(i} \d^{j)}_l
\right) \, , 
\ee
and the curvature scalar is
\be
R=   - \frac{1}{8} n(n-1)(n+2) \, .
\ee
The eigenvector to the eigenvalue 0 of the Ricci 
tensor is $g_{ij}$. All other eigenvalues equal $-n/4$,
 and 
the corresponding eigenvectors can be parametrized 
by the symmetric  traceless metrices, i.e.
the multiplicity of the eigenvalue $-n/4$ is $(n - 1)(n + 2)/2$.

\section{SUPERDETERMINANT}

We define the superdeterminant 
\be
     H = \det H_{AB} \, .
\ee
$H$  is a function of $g$, $\a$ and 
 $n$ which 
becomes zero for $\a = -1/n$, cf. sct. 3. 
We use eqs. (8) and (17) to look in more 
details for the explicit 
 value of $H$. The formal calculation for $n = 1$  leads to 
$$
H=H_{11} =G^{1111} =g^{11}  g^{11} + \a  g^{11}  g^{11}  
= (1+\a) g^{-2} \, .
$$
Multiplication of $g_{ij}$  with $\e$ 
gives $g \to \e^n g$,
 $ H_{AB} \to \e^{-2} H_{AB} $ and 
 $H \to \e^{-n(n+1)} H$. 
 So we get in an intermediate step
\be
     H = H_1 \, g^{-n-1}
\ee
where $H_1$ is 
the value of $H$ for $g= 1$. $H_1$ depends on $\a$ 
and $n$ only. To calculate $H_1$ we put 
 $g_{ij } = \d_{ij}$
 and get via $H_{ij} = \d_{ij} + \a$,
 $ H_{Ai} =  0$ 
for $A> n$, and $H_{AB} = \d_{AB}$  for 
$A, B > n$  finally
\be
 H_1=1+\a n \, .
\ee
This is in agreement with the $n = 1$-calculation.

\section{GRAVITY}

Now, we come to the main application: 
The action for gravity shall be expressed 
by the metric of superspace. We start from the metric
\be
     ds^2 = dt^2 - g_{ij} \, dx^i \, dx^j
\ee
$i,j =1, \dots n$  with positive definite $g_{ij}$  and $x^0 = t$.  
 We define the second fundamental form $K_{ij}$  by
\be
   K_{ij}  = \frac{1}{2} \, g_{ij,0} \, .
\ee
The  Einstein action for (35) is 
\be
 I = - \int \ {}^{\ast}R \ \frac{1}{2} \,  \sqrt g \, d^{n+1}x 
\ee
where $g  = \det g_{ij}$  and $ {}^{\ast}R$  is the 
 $(n + 1)$-dimensional curvature 
scalar for  (35). Indices at $K_{ij}$  will be 
shifted with $g_{ij}$, and $K = K^i_i$. With (36) we get
\be
     (K \sqrt g)_{,0} = (K_{,0} + K^2) \sqrt g \, . 
\ee
This divergence can  be added to the 
integrand of (37) without changing the 
field equations.  It serves to cancel the term $K_{,0}$
 of $I$. So we get
\be
 I =  \int \   \frac{1}{2} \left( 
 K^{ij} K_{ij} - K^2 + R 
 \right) 
  \sqrt g \, d^{n+1}x 
\ee
where $R$ is the $n$-dimensional curvature 
scalar for $g_{ij}$.  We make now the ansatz for the kinetic energy
\be
W =
   \frac{1}{2} G^{ijmp} K_{ij} K_{mp}
=
   \frac{1}{2} \left( 
 K^{ij} K_{ij} + \a  K^2  
 \right) \, .
\ee
Comparing (40) with (39) we see that 
for $\a = -1$  (surprisingly,  
this value does not depend on $n$) 
\be
 I =  \int \  \left( W +    \frac{R}{2} 
  \right) 
  \sqrt g \, d^{n+1}x 
\ee
holds. Because of $n \ge  2$  this value $\a$  
gives a regular superspace metric. (For 
 $n = 1$, eq. (37) is a divergence, 
and $\a = -1$ gives not an invertible superspace-metric.)

\medskip

Using the $\mu^{Aij}$  and 
the notations 
$z^A = \mu^{Aij} \, g_{ij } \, / \, 2$
 and $v^A =  dz^A/dt $ we get from  (40,41)
\be
 I =  \int \   \frac{1}{2} \left( 
 H_{AB} v^A v^B  + R(z^A)   \right)   \sqrt g \, d^{n+1}x 
\ee
i.e., the action has the classical form of  kinetic plus 
potential energy. The signature of the metric 
 $H_{AB}$  is $S = 1$. This can be seen from eqs. (18,24).

\section{CONCLUSION}

In eq. (42), Einstein gravity is given in a 
form to allow canonical quantization:
 The momentum $v^A$ is  replaced
 by   $- i \pa / \pa z^A $ \   ($\hbar =  1$), 
and then 
the Wheeler - DeWitt equation for the world 
function $\psi(z^A)$  appears as Hamiltonian constraint 
in form of  a wave equation:
\be
\left(
 \Box - R(z^A)
\right) \psi
=0 \, .
\ee
After early attempts in [1],  the 
Wheeler - DeWitt equation 
has often been discussed, especially for 
cosmology, see e.g. [2-5].  Besides curvature, 
matter fields can be inserted as potential, too. It
 is remarkable  that exactly for Lorentz and 
for Euclidean signatures in (35) (positive and
 negative definite $g_{ij}$  resp.) 
the usual D'Alembert operator  ($S = 1$) in (43) appears. For other 
signatures in (35), (43) has at least two timelike axes.

\bigskip

\noindent 
The author gratefully acknowledges stimulating 
discussions with U. Bleyer,  D.-E. Liebscher 
and A. I. Zhuk before and with 
J. Kijowski and P. Michor during the conference.

\section*{REFERENCES}

\noindent 
[1]  R. Arnowitt, S. Deser,  
C. Misner, 1962 in: E. Witten, Gravitation, An 
Introduction to current research, New York.

\noindent 
[2] U. Bleyer, D.-E. Liebscher, H.-J.
 Schmidt, A.I. Zhuk, PRE-ZIAP 89-11.

\noindent 
[3] G. Gibbons, S. Hawking, 
 J. Stewart, Nucl. Phys. B 281 (1987), 736.

\noindent 
[4]  L. P. Grishchuk, Yu. V. Sidorov, p. 700 in: 
Proc. 4. Sem. Quantum Gravity Moscow,
WSPC Singapore 1988, Ed. M. A. Markov.

\noindent 
[5]  J. Halliwell, S. Hawking, p. 509 in: Proc. 3. Sem.
 Quantum Gravity Moscow, WSPC
Singapore 1985, Ed. M. A. Markov.

\bigskip

\noindent
{\it  Received September 22, 1989}

\bigskip

\noindent 
  In this reprint 
we removed only obvious misprints of the original, which
was published in Proc. Conf. Brno August 27 - September 2, 1989: 
Differential    Geometry and its  Applications, Eds.: J.
Janyska, D. Krupka, World Scientific PC Singapore 1990, pages 
405-411.

\bigskip

\noindent 
  Author's address  that time: 
 H.-J. Schmidt,  Zentralinstitut f\"ur  Astrophysik der 
Akademie der Wissenschaften, 
DDR-1591 Potsdam, R.-Luxemburg-Str. 17a
\end{document}